# GNO SOLAR NEUTRINO OBSERVATIONS: RESULTS FOR GNO I

GNO COLLABORATION

M. Altmann[a], M. Balata[b], P. Belli[c], E. Bellotti[d], R. Bernabei[c], E. Burkert[e], C. Cattadori[d], G. Cerichelli[f], M. Chiarini[f], M. Cribier[g], S. d'Angelo[c], G. Del Re[f], K. H. Ebert[h], F. v. Feilitzsch[a], N. Ferrari[b], W. Hampel[e], J. Handt[e], E. Henrich[h], G. Heusser[e], J. Kiko[e], T. Kirsten[e], T. Lachenmaier[a], J. Lanfranchi[a], M. Laubenstein[b], D. Motta[d], W. Rau[e], H. Richter[e], S. +Wänninger[a], M. Wojcik[i], L. Zanotti[d]

[a] Physik Department E15, Technische Universität München (TUM), James-Franck Straße, D-85748 Garching, Germany [1]
[b] INFN, Laboratori Nazionali del Gran Sasso (LNGS), S.S. 17/bis Km 18+910, I-67010 l'Aquila, Italy [2]
[c] Dipartimento di Fisica, Universitá di Roma 'Tor Vergata' e INFN, Sezione di Roma II, Via della Ricerca Scientifica, I-00133 Roma, Italy [2]
[d] Dipartimento di Fisica, Universitá di Milano 'La Bicocca' e INFN, Sezione di Milano, Via Emanueli, I-20126 Milano, Italy [2]
[e] Max-Planck-Institut für Kernphysik (MPIK), P.O.B. 103980, D-69029 Heidelberg, Germany [3,4]
[f] Dipartimento di Ingegneria Chimica e Materiali, Universitá dell'Aquila, Località Monteluco di Roio, l'Aquila, Italy [2]
[g] DAPNIA/Service de Physique des Particules, CE Saclay, F-91191 Gif-sur-Yvette Cedex, France
[h] Institut für Technische Chemie, Forschungszentrum Karlsruhe (FZK), P.O.B. 3640, D-76021 Karlsruhe, Germany
[i] Instytut Fizyki, Uniwersytet Jagiellonski, ul. Reymonta 4, PL-30059 Kraków, Poland.

[1] This work has been supported by the German Bundesministerium für Bildung und Forschung (BMBF).
[2] This work has been supported by Istituto Nazionale di Fisica Nucleare (INFN), Italy.
[3] This work has been generously supported by the Alfried Krupp von Bohlen und Halbach-Foundation, Germany.
[4] This work has been supported by the German Bundesministerium für Bildung und Forschung ( BMBF)

submitted to Physics Letters B
June 2000

**Abstract**

We report the first GNO solar neutrino results for the measuring period GNO I, solar exposure time May 20, 1998 till January 12, 2000. In the present analysis, counting results for solar runs SR1 - SR19 were used till April 4, 2000. With counting completed for all but the last 3 runs (SR17 - SR19), the GNO I result is [65.8 ± $^{10.2}_{9.6}$ (stat.) ± $^{3.4}_{3.6}$ (syst.)] SNU (1σ) or [65.8 ± $^{10.7}_{10.2}$ (incl. syst.)] SNU (1σ) with errors combined. This may be compared to the result for Gallex (I-IV), which is [77.5 ± $^{7.6}_{7.8}$ (incl. syst.)] SNU (1σ). A combined result from both GNO I and Gallex (I-IV) together is [74.1 ± $^{6.7}_{6.8}$ (incl. syst.)] SNU (1σ).

## 1. Introduction

GNO (Gallium Neutrino Observatory) is the successor project of the Gallex solar neutrino experiment at the Gran Sasso Underground Laboratories (LNGS). The latter has been recording solar neutrinos with energies above 233 keV via the inverse beta decay reaction $^{71}Ga\ (\nu_e,e^-)^{71}Ge$ in a 100-ton gallium chloride target (containing 30.3 tons of gallium) between May 1991 and January 1997 [1-4]. After some further verification- and calibration experiments [5,6], the experimental program of Gallex was completed in fall 1997.

Without a funded new project, the gallium detector would have had to be dismantled, while the following motivations existed to continue the gallium observations (with experimental improvements and possibly with enlargement of the gallium mass):

*(i)* Gallex has observed a solar neutrino induced signal that is sufficient to account for the low-energy pp-neutrinos from the primary hydrogen fusion reaction in the solar core that are unavoidably associated with the production of the solar luminosity. However, in assigning the measured signal to pp-neutrinos (fusion chain 'PPI'), little or no signal is left to account for the other neutrino fluxes also expected from the Standard Solar Model (SSM) [7], in particular for $^{7}$Be neutrinos. This constitutes a real conflict (the '$^{7}$Be-neutrino problem' [2, 8-10]) since $^{7}$Be is the seed for the production of the $^{8}$B neutrinos that are actually observed to arrive on Earth in the Homestake and Superkamiokande experiments (even though only at about half of the expected strength).

While solar model modifications may affect the weak $^{8}$B neutrino branch, a (non-physical) 'zero-cross section' solar model can define an *absolute* minimum neutrino flux by setting *ad hoc* (and against better knowledge) all cross sections for PPII, PPIII, and for the CNO-cycle to zero, while the solar luminosity is sustained. For this one expects, free of assumptions, ≈ 80 SNU [11] (or ≈ 88 SNU if the CNO neutrinos were preserved). The Gallex result is 77.5 ± 8 SNU [4]. Suppose the experimental errors could be further reduced. In this case it could become possible to exclude standard (massless) neutrinos without any reference to solar models, at a confidence level > 95 %. To this end, improved statistics and a further reduction of systematic errors are required [12].

*(ii)* An intense program for a full spectroscopic exploration of solar neutrinos is presently under way [13]. However, without a running gallium detector, the majority (93 %) of solar neutrinos would remain unobserved. The determination of the charged current reaction rate for pp-neutrinos measured with high precision (≈ 5 %) by inverse beta decay on gallium will remain without competition for many years to come. Lack of this basic data would have a very negative impact also on the interpretation of the results from the other presently running solar neutrino experiments (Superkamiokande, SNO) as well as from forthcoming second generation solar neutrino experiments (Borexino, which is expected to start data taking in less than two years from now; LENS, which is in the R&D phase; and others).

*(iii)* Among the handful of neutrino mass and mixing parameter sets that are indicated to be allowed by the available experimental evidence, at least two can be singled out by time variations of the observed neutrino signals, particularly in GNO (seasonal and even day/night variation) [14-16].

After favorable review of the GNO proposal [12], the Gallium Neutrino Observatory (GNO) was approved and supported by INFN at LNGS in 1998. The project is intended to provide a long time record of low energy solar neutrinos, to determine the bulk production rate with an accuracy of 5 SNU, and to closely examine the time constancy of the pp-neutrino flux during a whole solar cycle with a sensitivity of ≈ 15 %.

After a major overhaul and modernization of the existing experimental set-up (which had been in continuous operation since 1990), GNO solar observations started in May 1998 [17,18]. Now we report the results based on the first 19 months of observation.

The plan of this article is as follows: In sect. 2 we give a short description of the experimental aspects that differ in GNO as compared to Gallex. In sect. 3 we present the first set of GNO data ('GNO I'). They were taken in 19 (about monthly) runs during the exposure time period May 20, 1998 - January 12, 2000. In sect. 4 we discuss the results of GNO I and the internal consistency of GNO and Gallex data combined (sect. 4.1). The possibility of time variations of the neutrino signal is addressed in sect. 4.2. Finally, we comment on the ongoing measurements and on future prospects of the GNO project (sect. 5). This includes in particular the development of alternative $^{71}Ge$ counting methods.

## 2. Experimental

*Target, runs, and extractions*

The basic procedures in performing GNO extraction runs have been similar as in Gallex [1-4,19]. Table 1 summarizes the GNO run characteristics. The operating period defined as GNO I covers runs A001 to A022 (2$^{nd}$ column). It comprises 19 solar runs that are consecutively labeled SR1 - SR19 (1$^{st}$ column). Runs A001 and A002 were initial test runs aimed to condition the system, run A012 was lost by operational failure. All other runs were successful. Among the 19 solar runs, 14 were exposed for 4 weeks, 4 for 5 weeks, and 1 for 6 weeks (5$^{th}$ column). During run A021, 164 kg of pure HCl has been added to the gallium chloride target in order to compensate for the cumulated HCl losses that occur during nitrogen purge in every run ('re-acidification').

Ge extraction yields were monitored with added Ge carriers (≈ 1 mg per run). $^{70}Ge$, $^{72}Ge$, $^{74}Ge$, and $^{76}Ge$ were used alternatingly (Table 1, 6$^{th}$ column). The Ge recovery yields are listed in column 7 of Table 1. The yield refers to the percentage of initially added germanium that actually ended up in the counter. The complement of 100 % is the sum of the Ge that escaped extraction from the target and of the unavoidable losses that occur during the conversion of $GeCl_4$ in $GeH_4$ (germane) and in the counter filling procedure. Yields range from 91.5 % to 97.9 %, with a mean of 95.3 %.

*Electronics and data acquisition*

Electronics, especially preamplifiers, pulse shape recording and data acquisition in GNO have been redesigned and implemented to replace and modernize the Gallex system that was designed > 10 years ago. The preamplifiers now have 300 MHz bandwidth for improved resolution in the analysis of the fast part of the pulses. The new system avoids multiplexing of the 16 lines by application of 32 individual fast transient digitizers ('TDF', 0.2 ns per channel for 400 ns). The amplification factors for L- and K pulses are separately optimized (28x and 4x, respectively). In addition to this high time resolution recording, the complete pulse is registered with (multiplexed) slow transient digitizers ('TDS', 400 ns/channel for 800 μs).

The new data acquisition system has significantly reduced the cross talk of digital noise with the analogue signals and thus improved the pulse shape discrimination potential between (mostly slow) background pulses and fast $^{71}$Ge decay pulses. Fig. 1 illustrates the GNO pulse recording quality for two genuine $^{71}$Ge pulses (one L, one K) and for two background pulses that are not trivial to identify as such because they have accidentally deposited the same energies in the counter as $^{71}$Ge decay pulses could have.

The new GNO data acquisition system has been designed for maximum flexibility. One of two 'Digital Alpha 1000' server units operating on Digital Unix performs the acquisition, while the other is a warm-backup that is also used for testing of system upgrades without disturbing the continuous data taking. The computers are linked with the counting system in the Faraday cage via high-speed optical links. The real time status of counting is regularly checked via remote access using a user-friendly software. Data storage is on a 20 Gb system of 6 hard discs.

A major improvement has been the implementation of a tunable power X-ray calibration source. It has replaced the radioactive $^{153}$Gd/Ce source that was formerly used for the energy calibration of the counters. Remote control of a system with two stepping motors can position the source to the desired counter position inside the shield without need to open the counting tank. This helps to reduce the systematic error due to calibration operations and allows more frequent calibrations with a reduced risk of disturbances. X-ray calibration is in use since start of counting for run SR13.

*Counting and pulse shape analysis*

All GNO measurements are performed in 'passive' counting positions (within a copper block). These positions have lower background and the diagnostic power of the 'active' positions within a NaI well veto counter is no longer required in normal runs.

In spite of the mean life of $^{71}$Ge being only 16.49 d, counting of a run lasts regularly for 5-6 months in order to fully characterize the counter background. In the present set of data, counting for runs SR17-19 is not yet completed (Table 1, last column) but sufficiently progressed to include these runs in the first analysis.

Counting efficiencies for 10 of the 13 counters used in GNO runs (see Table 1) have been derived by means of Monte Carlo (MC) simulations of $^{71}$Ge decays in the counters. For the other three counters (Fe039, Fe043 and SC135) efficiencies have been directly measured with real $^{71}$Ge activity at the end of GALLEX in their actual counting configuration. Comparison of measured efficiencies with those obtained by the MC method show agreement. Furthermore, they reveal that the errors of the calculated values depend mainly on the errors of the two most important single input data in the MC calculation, namely the counter volume efficiency and the gas amplification curve.

The K- and L rise-time acceptance window efficiencies were cross calibrated with a $^{153}$Gd/Ce source. For this, two prototypes for the major counters types, counters Fe43 and Si119, were doped with $^{71}$Ge activity (order 1 Bq).

At the same occasion, the L- and K peak efficiencies were found to be the same as formerly deduced in Gallex, within better than 1 %. The counter efficiencies for all counters used in GNO are quoted in Table 1, columns 9 and 10, separately for K- and L peak acceptance windows. Together, they range from 61-72 % for the various counters, with a mean of 67 % (≈ 31 % L + ≈ 36 % K).

Pulses that have passed the energy cut are next subjected to a rise time cut. Genuine $^{71}$Ge decays give a characteristic signal on the TDF, as they produce a point-like ionization in the counting gas: the recorded pulses are fast compared to most background events (see Fig.1). For the pulse shape discrimination, we use a one parameter cut. We consider the rise-time (RT) from 8 % to 60 % of the pulse amplitude. RT is required to be less than about 40-45 ns (depending on the specific counter and gas mixture).

The cut has a high efficiency for $^{71}$Ge signals (≥ 95 %) and rejects most of the background. The remaining background in the acceptance windows must then be rejected on the basis of the time momenta of the pulses (maximum likelihood analysis, see below).

## 3. GNO I - Results

As mentioned, only passive counting positions have been used for GNO I. A random sampling of counter types was applied for the measurements (see Table 1, 8$^{th}$ column). For the selection of the $^{71}$Ge events we have subjected the counting data to our standard energy-, rise time- and radon cuts and to the subsequent maximum likelihood analysis.

The total GNO I exposure time has been 574 days. The individual run results for the net solar production rates of $^{71}$Ge (based on the counts in the K and L energy- and rise time windows) are plotted in Figure 2 and listed in Table 2, *after* the subtraction of 4.25 SNU for side reactions and of 0.96 SNU to account for the < 100 % efficiency of our cut for rejection of radon induced counts (see below).

In spite of the low statistical significance of any single runs, the combined analysis of 19 runs yields a rather significant result (last line of Table 2).

The result of the maximum likelihood analysis identifies a total of 90 decaying $^{71}$Ge atoms that were observed during 574 days of neutrino exposure. 83 of them (or 4.4 per run) are due to solar neutrinos. The mean $^{71}$Ge count rate during the first half-life of $^{71}$Ge decay (11.43 d) per run and counter is ≈ 0.23 counts per day of 'on-time' in counting. This may be compared with the time *independent* counter background in the acceptance windows. The latter is in average as low as ≈ 0.07 counts per day (Table 3).

The accurate determination of the background levels has benefited from our routinely long counting time (≈ 6 months), so most of the time independent background dominates while the $^{71}$Ge has long been decayed. Relevant background results for the counters used in GNO I are compiled in Table 3.

All run results have been corrected for $^{71}$Ge produced in side reactions, as explored in Gallex [20] and specified in Table 4. These figures are applicable since, relative to Gallex, conditions have not changed, except for a detail mentioned in the Table caption.

The systematic errors are specified in Table 5. The total systematic 1σ error amounts to ≈ 5 %. It is still dominated by the error of the counter efficiencies.

The distribution of the individual GNO I results and their statistical fluctuations follow a Gaussian distribution (see sect. 4.1). The mean-life fit for $\tau_{71} = \tau(^{71}\text{Ge})$ from all GNO I runs is $\tau_{71} = 22.5 \pm ^{6.8}_{5.5}$ d (1σ), compatible with the known value, 16.49 d [22].

After subtraction of 5.21 ± 0.96 SNU for side reactions and for the Rn-cut inefficiency (Table 4), the net result for GNO I is $65.8 \pm ^{10.7}_{10.2}$ SNU (1σ). This may be compared with the Gallex result, which is $77.5 \pm ^{7.6}_{7.8}$ SNU.

## 4. Discussion

### 4.1 GNO results and context with the available Gallex data

The scatter among the GNO I single run results agrees with the hypothesis of a constant production rate. This has been assured by application of the maximum likelihood ratio test (see [2] for details). The resulting goodness of fit confidence level is 36 %.

The GNO I result is within 1σ of the Gallex result. The relevant data are listed in Table 6 and the individual run results are plotted together for GNO and Gallex in Figure 3.

In order to judge on the internal consistency of GNO- and Gallex data under the premise of a signal constant in time, we have done an analysis in which the periods GNO I and Gallex IV, III, II, and I with their statistical errors are compared to the result from all GNO- and Gallex runs taken together (assumed to be the 'true value' in the $\chi^2$-calculation). In the hypothesis of a production rate that is constant over the entire data taking period, the $\chi^2$ is 10.6 (4 d.o.f.), corresponding to a probability of 3.2 %. Though this probability is low, it is higher than for Gallex alone and in any case consistent with the proposition that the gallium signal has been constant in time from 1991 through early 2000 (see also the related discussion in [4]).

The combined GNO + Gallex result after 84 solar runs is $74.1 \pm ^{6.7}_{6.8}$ SNU. It agrees well with the result of the SAGE experiment, 67.2 ± 8 SNU ($1\sigma$) [23]. The general implications of the fact that our result is substantially below the predictions of the various standard solar models (see sect. 1; range 120-140 SNU) have been discussed [4,9] and are not repeated here.

### 4.2 Possible time variations of the incident neutrino fluxes

The finding that the solar data are consistent with a production rate constant in time does not invalidate other hypotheses that might give similar or even better time dependent (linear, periodic, or other) fits. There have been analyses of the GALLEX data in terms of possible (anti/)correlation with the seasonal Sun - Earth distance variation and caused by neutrino oscillations [14,15]. Therefore, we have grouped the results for the 84 solar neutrino runs of GNO I and Gallex in 6 about equally populated bins of similar heliocentric distance, d. Figure 4 is the plot of the joint results for each of these bins in SNU vs. < d > in A.U.

No obvious feature is apparent, the fit relative to the expectation for a solar neutrino output constant in time (solid line in Figure 4) yields a confidence level of 13 % (5 d.o.f.). We remark that certain features reported in the above mentioned earlier analyses of the GALLEX data are rather unstable against added new input data and against model parameters. We are presently preparing a respective study that includes the new data.

## 5. Future plans

*Development of new counters*

In an alternative approach towards a further reduction of the counting errors, we are developing a machined counter that is more uniform in shape than the present suprasile quartz counters, yet competitive with the latter in terms of radiopurity (low background). The new counters may be produced in large numbers for one-time use, thus allowing active internal calibration with $^{71}$Ge after completion of the solar run counting.

Progress has also been made in our efforts to develop a low temperature calorimeter with the potential to increase the $^{71}$Ge counting efficiency from ≈ 70 % to near 100 % [24]. The Ge to be counted is thermally deposited onto a sapphire absorber crystal in which the energy deposit from a decay event leads to a temperature increase that is measured with a superconducting phase transition thermometer. The resistance change is read out with a superconducting quantum interference device (SQID).

We have now optimized the deposition technique for Ge and developed a dedicated detector set-up suitable for application in GNO. We have achieved a germanium deposition efficiency of 95 %, with potential for further improvement. Efficiency determination and control are done with several independent methods, the accuracy is better than 1 %.

In our counting device, two low temperature calorimeters of dimensions 1 x 2 x 0.1 cm$^3$ face each other, with the Ge deposited as a thin layer in between. This provides the required 4π angular acceptance for detection of germanium K- and L-capture decays.

Detailed Monte Carlo simulations, in addition to estimating the overall detection efficiency, allowed to associate the structures seen in the measured spectrum with the different de-excitation channels that follow the electron capture decay of $^{71}$Ge. Forthcoming issues are the demonstration of a sufficiently stable long-term operation of the detectors in a low-background environment and material selection according to radiopurity requirements.

*Present activities*

GNO data taking continues. At the time of this writing, we have already performed five more solar runs (SR20 - SR24, Jan 14 - May 30, 2000) and two blank runs, they are presently being counted. Other ongoing activities are: *(i)* a more sophisticated pulse shape analysis based on fitting the whole pulse rather than rise-time only, *(ii)* the quantitative verification of the individual Ge extraction yields by mass spectrometry, and *(iii)* the re-determination of the radon cut efficiency in a dedicated experiment with a minimally Ra-doped counter (counting since ≈ 1 yr).

*Outlook*

In the long run, up-scaling to 100 tons of gallium is envisioned for a reduction of the total error to ≈ 4 SNU. It is planned to increase the target size in steps from 30 tons of Ga to 60 t (GNO 60) and then to 100 t (GNO 100). To make this pay, the systematic error is also reduced further. The past and anticipated future development of the experimental errors in GNO is shown in Fig. 5, contingent to additional gallium.

## Acknowledgements

We thank the Gallex Collaboration for the permission to access the Gallex raw data for a joint analysis in combination with GNO data. We are obliged to A. Bettini, director of LNGS, for his competent and productive support and advice. We also want to acknowledge the LNGS staff, especially A. Falgiani and S. Nisi of the LNGS chemistry group. The skilful help of R. Lackner (MPIK) in the construction of the X-ray calibration unit is thankfully acknowledged.

**References:**

[1] Gallex Collaboration, P.Anselmann et. al., Phys. Lett. B285 (1992) 376

[2] Gallex Collaboration, P.Anselmann et. al., Phys. Lett. B357 (1995) 237

[3] Gallex Collaboration, W. Hampel et. al., Phys. Lett. B388 (1996) 384

[4] Gallex Collaboration, W. Hampel et. al., Phys. Lett. B447 (1999) 127

[5] Gallex Collaboration, W. Hampel et. al., Phys. Lett. B420 (1998) 114

[6] Gallex Collaboration, W. Hampel et. al., Phys. Lett. B436 (1998) 158

[7] J. N. Bahcall, S. Basu and M. Pinsonneault, Phys. Lett. B433 (1998) 1

[8] X. Shi, D. Schramm and D. Dearborn, Phys. Rev. D50 (1994) 2414

[9] T. A. Kirsten, Ann. N.Y. Acad. Sci. 759 (1995) 1

[10] J. N. Bahcall, P. I. Krastev and A. Smirnov, Phys. Rev. D58 (1998) 096016

[11] J. N. Bahcall, Phys. Rev. C56 (1997) 3391

[12] Proposal for a permanent Gallium Neutrino Observatory (GNO) at Laboratori Nazionali del Gran Sasso. INFN/AE-96-27, also in: LNGS Annual Report (1995) 41-84

[13] T. A. Kirsten, Rev. Mod. Phys. 71 (1999) 1213

[14] V. Berezinsky, G. Fiorentini and M. Lissia, Astroparticle Physics 12 (2000) 299

[15] G. L. Fogli et al., Phys. Rev. D61 (2000) 073009

[16] S. Wänninger et al., Phys. Rev. Lett. 83 (1999) 1088

[17] E. Bellotti, Proceed. 4th Intern. Solar Neutrino Conference, edited by W. Hampel, (Max-Planck Institut für Kernphysik, Heidelberg, Germany) (1997) 173

[18] T. A. Kirsten, Nucl. Phys. B. (Proceed. Suppl.) 77 (1999) 26

[19] E. Henrich and K. H. Ebert, Angew. Chemie Int. Ed. (Engl.) 31 (1992) 1283

[20] Gallex Collaboration, P.Anselmann et. al., Phys. Lett. B314 (1993) 445

[21] MACRO Collaboration, Astroparticle Physics 10 (1999) 11

[22] W. Hampel and L. P. Remsberg, Phys. Rev. C31 (1985) 677

[23] Sage Collaboration, J.N. Abdurashitov et al., Phys. Rev. Lett. 83 (1999) 4686

[24] M. Altmann et al., Nucl. Phys. B (Proc. Suppl.) 70 (1999) 374.

**Table 1:** GNO I - Run Characteristics

| Type[a] | Run # | Exposure time Start | End | Duration [d] | Carrier[b] | Ge yield [c] [%] | Label [d] | Counting effic. [e] L [%] | K [%] | Counting time [d] |
|---|---|---|---|---|---|---|---|---|---|---|
| | A001 | (02/97) | 23/04/98 | - | 76 | | | | | |
| | A002 | 23/04/98 | 20/05/98 | 27 | 70 | | | | | |
| SR1 | A003 | 20/05/98 | 17/06/98 | 28 | 72 | 96.1 | SC 138 | 32.6 | 38.1 | 179.6 |
| SR2 | A004 | 17/06/98 | 22/07/98 | 35 | 74 | 93.5 | Fe118 | 27.4 | 33.7 | 173.5 |
| SR3 | A005 | 22/07/98 | 26/08/98 | 35 | 76 | 95.1 | Si114 | 31.0 | 33.5 | 162.3 |
| SR4 | A006 | 26/08/98 | 23/09/98 | 28 | 70 | 97.9 | Si 113 | 30.4 | 34.6 | 138.3 |
| SR5 | A007 | 23/09/98 | 21/10/98 | 28 | 72 | 94.6 | FC093 | 32.4 | 38.2 | 137.7 |
| SR6 | A008 | 21/10/98 | 18/11/98 | 28 | 74 | 94.5 | Si108 | 30.3 | 34.8 | 166.5 |
| SR7 | A009 | 18/11/98 | 16/12/98 | 28 | 76 | 94.4 | SC136 | 32.9 | 39.2 | 180.6 |
| SR8 | A010 | 16/12/98 | 13/01/99 | 28 | 70 | 96.8 | FC102 | 31.4 | 37.4 | 179.4 |
| SR9 | A011 | 13/01/99 | 10/02/99 | 28 | 72 | 95.8 | SC135 | 32.9 | 39.2 | 194.7 |
| - | A012 | 10/02/99 | 10/03/99 | 28 | 74 | | - | - | - | - |
| SR10 | A013 | 10/03/99 | 14/04/99 | 35 | 76 | 94.6 | SC139 | 32.5 | 38.7 | 187.4 |
| SR11 | A014 | 14/04/99 | 19/05/99 | 35 | 70 | 96.4 | Fe039 | 27.8 | 33.9 | 180.4 |
| SR12 | A015 | 19/05/99 | 16/06/99 | 28 | 72 | 95.7 | Fe043 | 28.5 | 34.9 | 179.4 |
| SR13 | A016 | 16/06/99 | 28/07/99 | 42 | 74 | 94.4 | SC136 | 32.9 | 39.2 | 165.8 |
| SR14 | A017 | 28/07/99 | 25/08/99 | 28 | 76 | 96.0 | FC093 | 31.0 | 36.7 | 167.6 |
| SR15 | A018 | 25/08/99 | 22/09/99 | 28 | 70 | 97.1 | FC102 | 31.4 | 37.4 | 165.6 |
| SR16 | A019 | 22/09/99 | 20/10/99 | 28 | 72 | 95.7 | Si113 | 29.7 | 34.9 | 165.4 |
| SR17 | A020 | 20/10/99 | 17/11/99 | 28 | 74 | 95.6 | SC139 | 32.5 | 38.7 | 137.5 |
| SR18 | A021 | 17/11/99 | 14/12/99 | 27 | 76 | 94.7 | Fe039 | 27.8 | 33.9 | 110.6 |
| SR19 | A022 | 14/12/99 | 12/01/00 | 29 | 70 | 91.5 | Si106 | 30.0 | 34.4 | 81.5 |

[a] SR = solar neutrino run.
[b] 70, 72, 74, 76 indicate the use of carrier solutions enriched in $^{70}$Ge, $^{72}$Ge, $^{74}$Ge, $^{76}$Ge, respectively.
[c] Integral tank-to-counter yield of Ge-carriers. The combined error assigned to the uncertainties of the yields and of the target mass is 2.2 %.
[d] Counters have either iron or silicon cathode. SC = silicon counter with shaped cathode, FC = iron counter with shaped cathode.
[e] Rise time cut included. The window efficiencies introduce a systematic error of 4.2 %.

**Table 2:** Results for individual solar neutrino runs in GNO I

| Run number | | Result [SNU] |
|---|---|---|
| SR1 | A003 | $71 \pm ^{45}_{36}$ |
| SR2 | A004 | $48 \pm ^{44}_{34}$ |
| SR3 | A005 | $97 \pm ^{63}_{51}$ |
| SR4 | A006 | $69 \pm ^{51}_{40}$ |
| SR5 | A007 | $-46 \pm ^{40}_{33}$ |
| SR6 | A008 | $45 \pm ^{50}_{38}$ |
| SR7 | A009 | $116 \pm ^{53}_{44}$ |
| SR8 | A010 | $-51 \pm ^{46}_{37}$ |
| SR9 | A011 | $126 \pm ^{56}_{47}$ |
| SR10 | A013 | $123 \pm ^{53}_{43}$ |
| SR11 | A014 | $53 \pm ^{41}_{30}$ |
| SR12 | A015 | $26 \pm ^{69}_{50}$ |
| SR13 | A016 | $97 \pm ^{53}_{43}$ |
| SR14 | A017 | $114 \pm ^{69}_{55}$ |
| SR15 | A018 | $46 \pm ^{38}_{28}$ |
| SR16 | A019 | $49 \pm ^{46}_{36}$ |
| SR17 | A020 | $33 \pm ^{32}_{34}$ |
| SR18 | A021 | $67 \pm ^{44}_{40}$ |
| SR19 | A022 | $79 \pm ^{57}_{41}$ |
| all: | L only | $80.0 \pm ^{17.5}_{16.2}$ |
| all: | K only | $57.2 \pm ^{12.4}_{11.4}$ |
| **all:** | **GNO I** | $\mathbf{65.8} \pm ^{\mathbf{10.2}}_{\mathbf{9.6}}$ |

All SNU-values shown are net solar production rates of $^{71}Ge$ after subtraction of 5.21 SNU for side reactions (4.25 SNU) and for the Rn cut inefficiency (0.96 SNU, see text). The quoted errors are statistic only.

**Table 3:** Background rates for counters used in GNO I (0.5 keV ≤ E ≤ 15 keV)

| Counter | Background rate [counts/11.43 d][a] | | |
|---|---|---|---|
| | L-window fast, > 0.5 keV | K-window fast | Integral all |
| SC135 | 0.22 | 0.005 | 2.8 |
| SC136 | 0.51 | 0.17 | 4.1 |
| SC138 | 0.19 | 0.09 | 3.6 |
| SC139 | 0.21 | 0.17 | 3.0 |
| Si108 | 0.29 | 0.29 | 3.7 |
| Si113 | 0.55 | 0.39 | 4.5 |
| Si114 | ≤ 1.4 | ≤ 0.63 | 6.3 |
| FC093 | 0.72 | 0.32 | 7.1 |
| FC102 | 0.99 | 0.26 | 5.9 |
| Fe039 | 0.18 | 0.09 | 5.7 |
| Fe043 | 1.7 | 0.98 | 10.1 |
| Fe118 | 0.19 | 0.31 | 5.3 |
| BACKGROUND-COUNTS PER DAY | | | |
| range | 0.02-0.15 | 0.00-0.09 | 0.2-0.9 |
| mean | 0.046 | 0.025 | 0.45 |
| Sum of fast K + L : 0.071 cpd | | | |

For the K- and L energy windows, the quoted rates apply to the fast background pulses that can mimic $^{71}$Ge pulses. The integral rate includes both, fast as well as slow pulses in the whole energy range 0.5 keV ≤ E ≤ 15 keV.

[a] 11.43 d = half-life of $^{71}$Ge

**Table 4:** Side reaction subtractions to be applied to solar neutrino runs

| | |
|---|---|
| Muon induced background [a,b] | 3.1 ± 0.6 SNU |
| Fast neutrons | 0.15 ± 0.1 SNU |
| $^{69}Ge$ produced by muons and $^{8}B$ neutrinos and falsely attributed to $^{71}Ge$ [a] | 1.0 ± 0.5 SNU |
| Subtotal | 4.25 ± 0.8 SNU |
| Rn-cut inefficiency [a] | 0.96 ± 0.53 SNU |
| **Total to subtract** | **5.21 ± 0.96 SNU** |

[a] For details how these values are derived, see [20].
[b] adjusted for new MACRO muon flux measurement at LNGS [21].

**Table 5:** Systematic errors in GNO I

| | [%] [a] | [SNU] |
|---|---|---|
| Counting efficiency including energy- and rise-time cuts | + 4.1 / – 4.4 | + 2.90 / – 3.09 |
| target size and chemical yield | ± 2.2 | ± 1.52 |
| side reactions subtraction error [b] | ± 1.4 | ± 0.96 |
| **Total systematic errors** | **+ 4.9 / – 5.1** | **+ 3.4 / – 3.6** |

[a] before subtractions (100 % = 71.0 SNU)
[b] see table 4, last line.

**Table 6:** Results from GNO and Gallex

| | **GNO I** | **Gallex (I-IV)** | **GNO + Gallex** |
|---|---|---|---|
| Time period | 05/20/98–01/12/00 | 05/14/91 – 01/23/97 [a] | 05/14/91– 01/13/2000 [b] |
| Net exposure days [d] | 574 | 1594 | 2168 (5.94 yrs) |
| Number of runs | 19 | 65 | 84 |
| L only [SNU] (stat.error only) | $80.0 \pm ^{17.5}_{16.2}$ | 74.4 ± 10 | 75.7 ± 8.6 |
| K only [SNU] (stat.error only) | $57.2 \pm ^{12.4}_{11.4}$ | 79.5 ± 8.2 | 73.1 ± 6.7 |
| Result (all) [SNU] | $65.8 \pm ^{10.2}_{9.6}$ stat. $\pm ^{3.4}_{3.6}$ syst. | 77.5 ± 6.2 stat. $\pm ^{4.3}_{4.7}$ syst. | 74.1 ± 5.4 stat. $\pm ^{4.0}_{4.2}$ syst. |
| **Result (all)** [SNU] [c] | **65.8** $\pm ^{\mathbf{10.7}}_{\mathbf{10.2}}$ incl. syst. | $77.5 \pm ^{7.6}_{7.8}$ incl. syst. [4] | **74.1** $\pm ^{\mathbf{6.7}}_{\mathbf{6.8}}$ incl. syst. |

[a] except periods of no recording: 5-8/92; 6-10/94; 11/95-2/96
[b] except periods of no recording: as before plus 2/97-5/98
[c] statistical and systematic errors combined in quadrature. Errors quoted are 1σ.

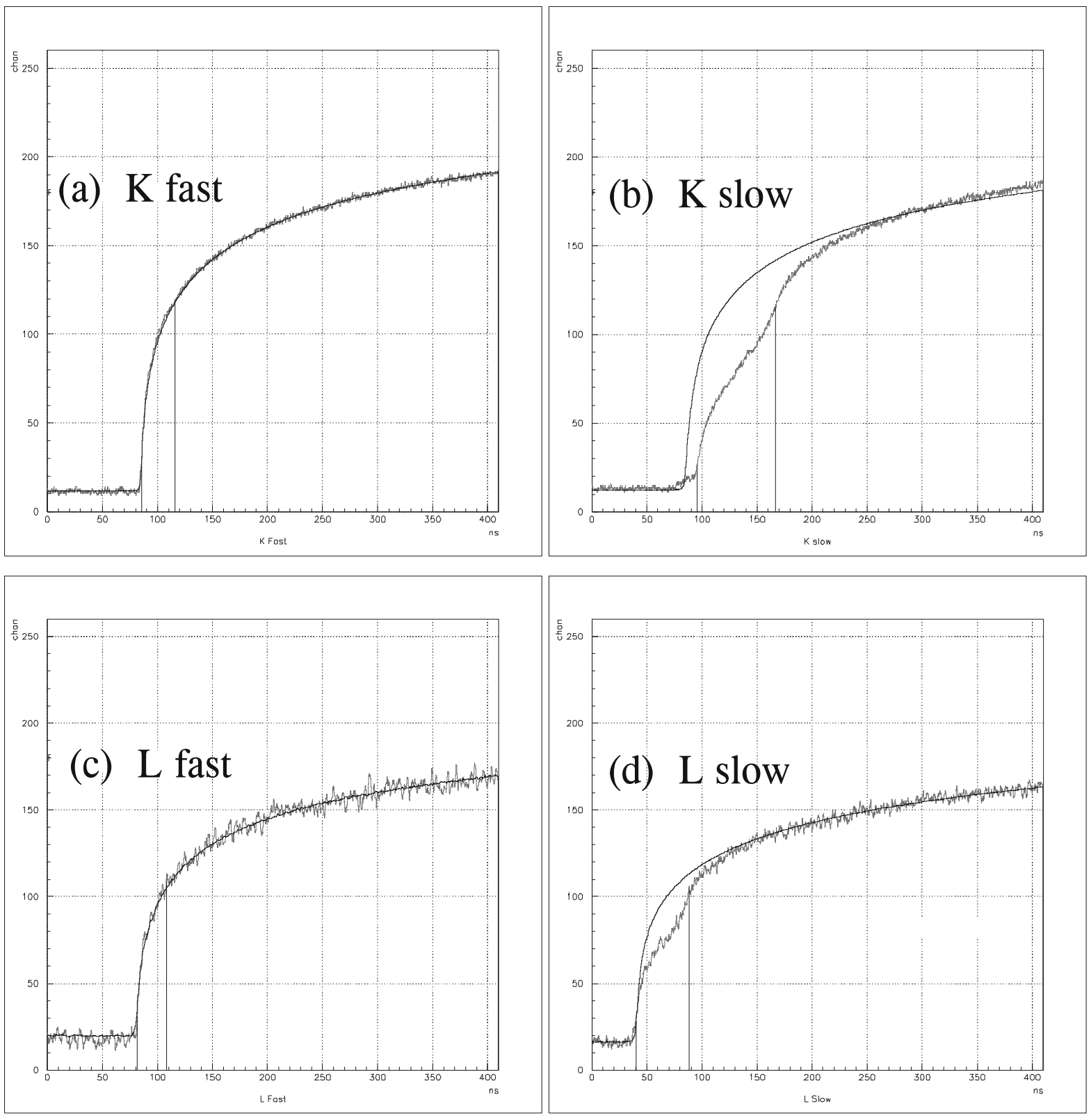

**Figure 1**: Examples of 4 pulses recorded by TDF's in GNO.

$^{71}$Ge electron captures (both from the K and L shells) produce fast pulses, while most background events generate slow pulses from diffuse ionization. *(a)* fast K pulse, ≈ 10.4 keV *(c)* fast L pulse ≈ 1.2 keV *(b)* and *(d)*: slow background pulses with energy deposition identical to K (L) $^{71}$Ge pulses (indicated by the solid reference lines for the expected averages for fast pulses from $^{71}$Ge decay).

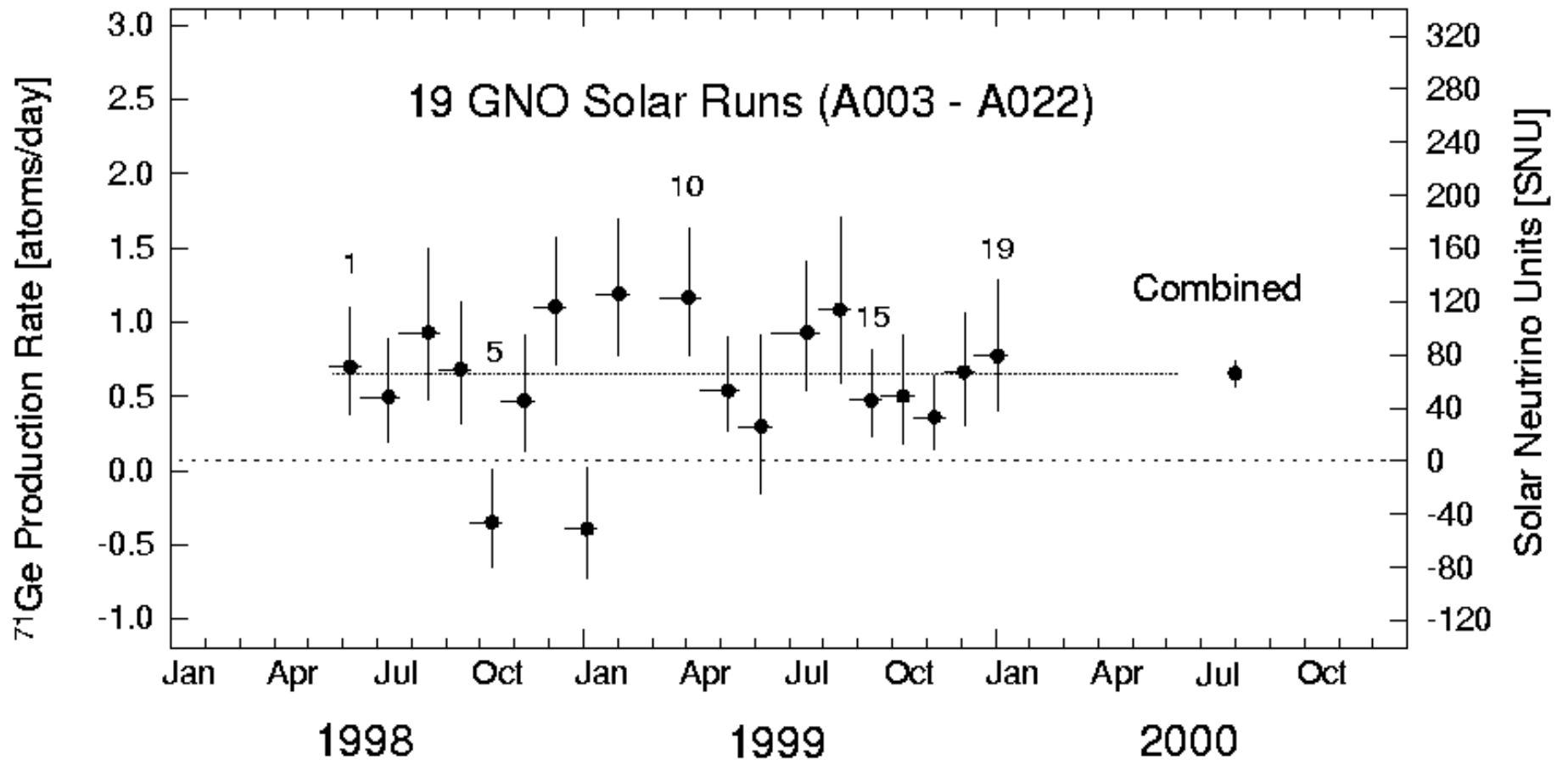

**Figure 2:** GNO single run results.

The left hand scale is the measured $^{71}$Ge production rate; the right hand scale is the net solar neutrino production rate (unit = SNU) after subtraction of side reaction contributions (see text). Error bars are ± 1σ, statistical only. To aid orientation, some SR numbers (Tables 1 and 2, 1st columns) are printed above the respective data points. Horizontal bars represent run duration; their asymmetry reflects the 'mean age' of the $^{71}$Ge atoms produced.

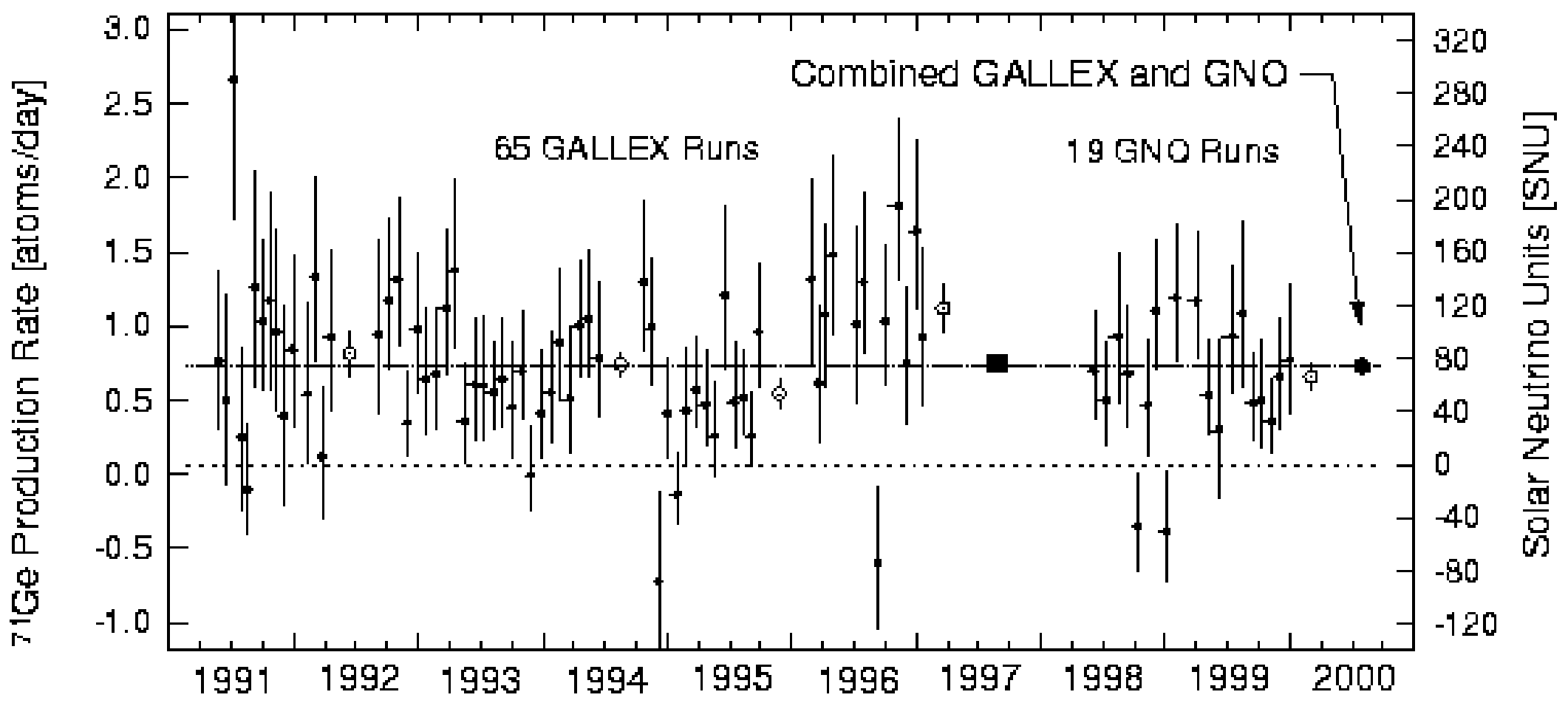

**Figure 3:** GNO and Gallex single run results.

The results for the 19 solar neutrino runs of GNO I are shown together with the earlier results from Gallex. The left hand scale is the measured $^{71}$Ge production rate; the right hand scale, the net solar neutrino production rate (SNU) after subtraction of side reaction contributions (see text). Error bars are ± 1σ, statistical only. Open dots represent group mean values; the bold square is for Gallex (all); the bold solid dot represents the global result for GNO + Gallex together (all 65 + 19 = 84 runs).

Some of the *individual* run results plotted for Gallex differ somewhat from what has been published by the Gallex collaboration [4]. This is due to some late refinements in the data evaluation and better knowledge of counter efficiencies. Notwithstanding, the overall result and error for Gallex has remained virtually unchanged. The publication of these details by Gallex is in preparation.

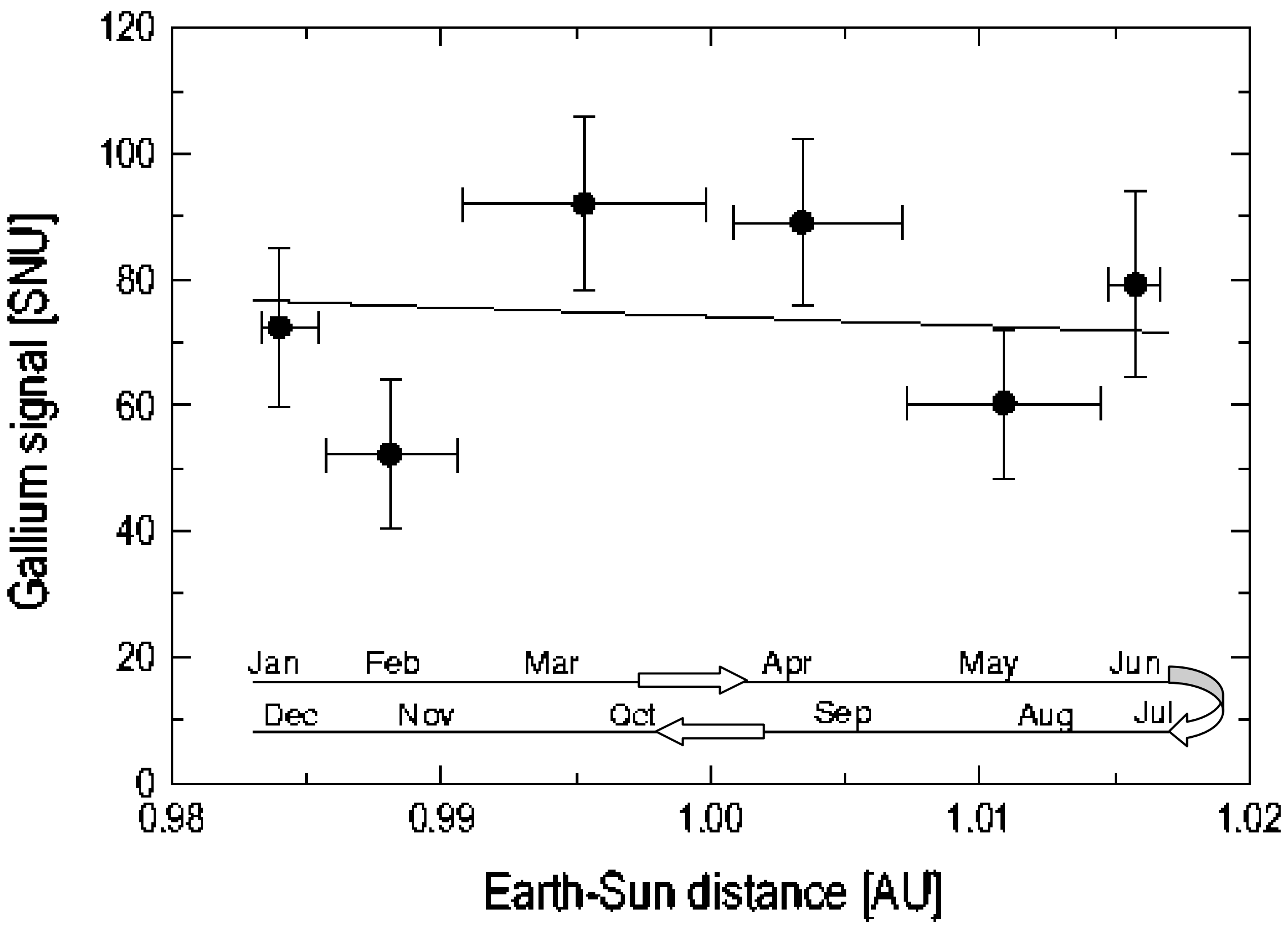

**Figure 4: Gallium signal vs. heliocentric distance of the Earth**

Approximate corresponding calendar months are indicated at the top of the abscissa. The solid line indicates the expected flux variation for purely geometrical ($1/d^2$) reasons. The mean value for d = 1 is set equal to 74.1 SNU.

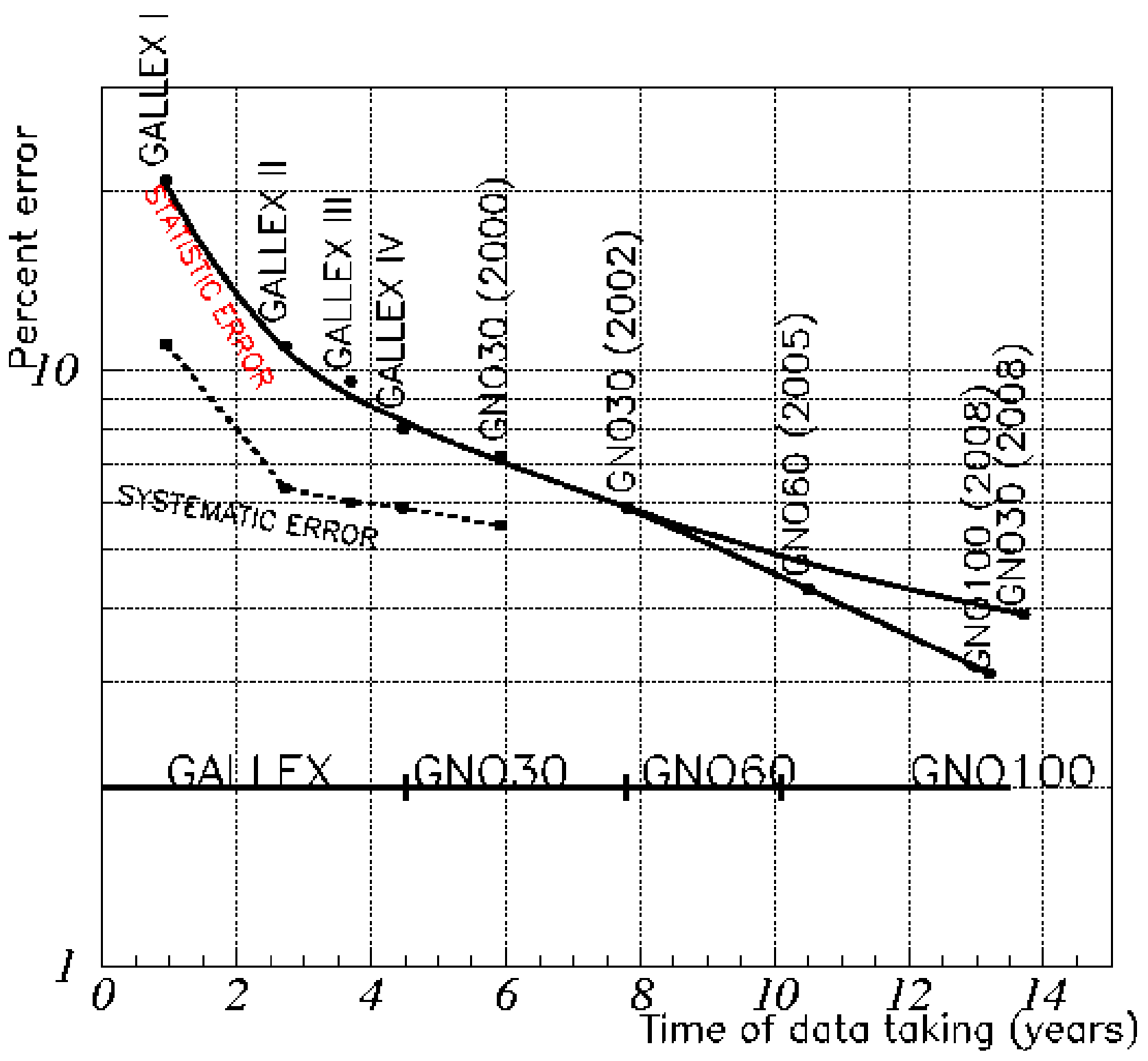

**Figure 5:** Time evolution of the errors in GNO

The time schedule of additional Ga acquisition is assumed: GNO30 = no additional Ga, GNO60 = 30 additional tons of Ga, GNO100 = further 40 tons of Ga. The development of systematic errors justifies target mass enlargements. Numbers in parentheses indicate the calendar year.